\documentclass{article}
\usepackage{glas}
\usepackage[english]{babel}
\usepackage{cite}
\usepackage{xspace}
\usepackage{epsf}
\epsfverbosetrue

\newcommand{\kt}{\mbox{$<\!\!k_T\!\!>\,$}}

\newcommand{\ppp}{\mbox{$<\!\!p_\perp\!\!>\,$}}
\newcommand{\jet}{{\,\mathrm{jet}}}
\newcommand{\meas}{{\,\mathrm{meas}}}
\newcommand{\yJB}{y_{\mathrm{JB}}}

\begin{document}
\begin{titlepage}{GLAS-PPE/2001--03}{April 2001}
\title{\bf Prompt Photons in Photoproduction at HERA}
\author{P. J. Bussey}
\vspace{-0.5cm}
\collaboration{for the ZEUS Collaboration}
\begin{abstract}
The production of prompt photons in photoproduction
reactions in the ZEUS experiment is discussed.  Cross sections 
for inclusive prompt photons and for prompt photons accompanied
by jets are compared with theory, and the latter are used to
determine the effective transverse momentum of the quarks in
the proton.
\end{abstract}
\conference{Paper presented at the 9th International Workshop\\on Deep Inelastic Scattering,\\Bologna, 27 April -- 1 May 2001}
\end{titlepage}

\section{Introduction.}
The study of parton interactions is frequently hindered by the 
fact that outgoing quarks and gluons cannot be observed directly. 
Photons, however, referred to as ``prompt photons'', can emerge as a
primary product of hard parton processes without the
hadronisation by which quarks and gluons form
jets.  In this way, they provide information 
that is relatively free from hadronisation
uncertainties.

This talk summarises two recent analyses by ZEUS at HERA\cite{Z2,Z3},
which present measurements of prompt photon cross sections in
photoproduction.

\section{Experimental method.}

Prompt photons are measured in the ZEUS Barrel Calorimeter.
The fine-grained, pointing geometry of the electromagnetic section 
allows photons to be distinguished from $\pi^0$ and $\eta^0$
backgrounds by first removing those candidates consisting of a broad
cluster of fired cells. A statistical subtraction is then performed as  
a function of the fraction of the energy in the highest-energy
cell.  An isolation condition removes further background, 
and reduces the contribution from photons radiated from final-state quarks.

The integrated $e^+p$ luminosity was $38.6\pm0.6$ pb$^{-1}$, with $E_e
= 27.5$ GeV and $E_p = 820 $ GeV.  The analysis employed energy-flow
objects which combine information from calorimeter cells and measured
tracks.  The incoming virtual photon energy was estimated using $\yJB$
= \mbox{$\sum(E - p_Z)/2E_e$,} summing over all energy-flow objects in
the event, each treated as equivalent to a massless particle.  After
correcting for energy losses, limits on $\yJB$ were applied,
corresponding to limits on the centre-of-mass $\gamma p$ energy $W$.
Events with a scattered beam positron in the calorimeter
were rejected both explicitly and by the upper limit on $\yJB$.

Jets were reconstructed, using energy-flow objects, by means of the
Lorentz-invariant $k_T$-clustering algorithm KTCLUS in the inclusive
mode.  The momenta of the objects comprising the jet were
summed to obtain the total jet-momentum vector.  An energy correction
was applied.   

After correction, the photons and jets were respectively required to have
$E_T^\gamma > 5$ GeV with $-0.7 < \eta^\gamma < 0.9$, and $E_T^{\jet}
> 5$ GeV  with $-1.5 < \eta^{\jet} < 1.8$, where $\eta$
is the laboratory pseudorapidity.  The fraction of the incoming photon
energy that takes part in the QCD subprocess was estimated as
$x_\gamma^\meas = \sum_{\gamma, {\jet}}\,(E-p_Z)/{2E_e\,\yJB},$
summing over the high-energy photon and the jet.


\begin{figure}[b!]
\centerline{\hspace*{4mm}\epsfig{file=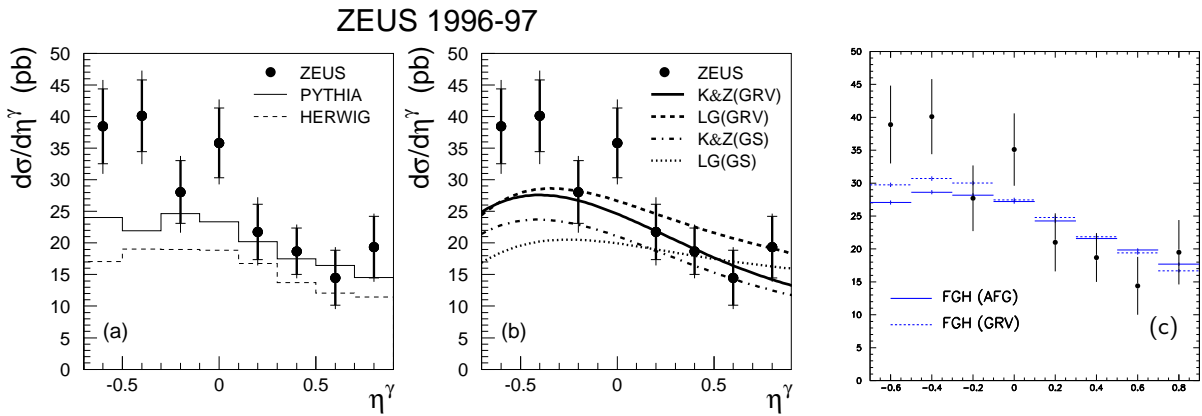,width=15.4cm,%
bbllx=0pt,bblly=380pt,bburx=480pt,bbury=560pt,clip=yes} 
\hspace*{-8mm}\raisebox{14mm}{{\small\sf(c)}}}
\caption{\small Inclusive prompt photon distribution in ZEUS for events in
the range $134 < W < 285$ GeV.  The data are compared with (a)
Hadronising Monte Carlos, (b) NLO calculations of Krawczyk and
Zembrzuski (K\&Z) and Gordon (LG), (c) the recent calculation of
FGH\protect\cite{FGH}.  Of the various photon parton distributions
tested, that of GRV appears preferred. }
\end{figure}

\section{Inclusive photoproduction of prompt photons.} 

Figure 1 shows the inclusive prompt photon cross section as a function
of $\eta^\gamma$.  The magnitude is approximately described by PYTHIA;
HERWIG is lower due to a smaller radiative contribution.
The NLO parton-level calculations give a fair description.  However at
negative $\eta^\gamma$, all the models are low.  
The shape of the $E_T^\gamma $
distribution is well described by each model.  There are important
contributions from the direct and resolved processes, and also from 
radiative processes.

The recent FGH calculation\cite{FGH}, still at the parton level,
contains the NLO terms in LG and also a box diagram term that is
included in K\&Z.  It is a slight improvement, but again fails to
account fully for the backward cross section.  Varying the effective
transverse momentum of the partons in the proton has been
investigated, and seems unable to generate the effect.  As shown below,
there is no significant discrepancy when a jet is included in
the cross section definition; it is therefore worth asking whether
the processes which give a very forward-backward event need special
theoretical treatment.  A recent calculation\cite{Metz} predicts a
BFKL contribution of a suitable magnitude, namely
$d\sigma/d\eta^\gamma \approx 8$ pb, but with no backward peaking.  It
will be of interest to see whether the CCFM approach\cite{Lonn} can
assist.

\begin{figure}[t]
\mbox{\epsfig{file=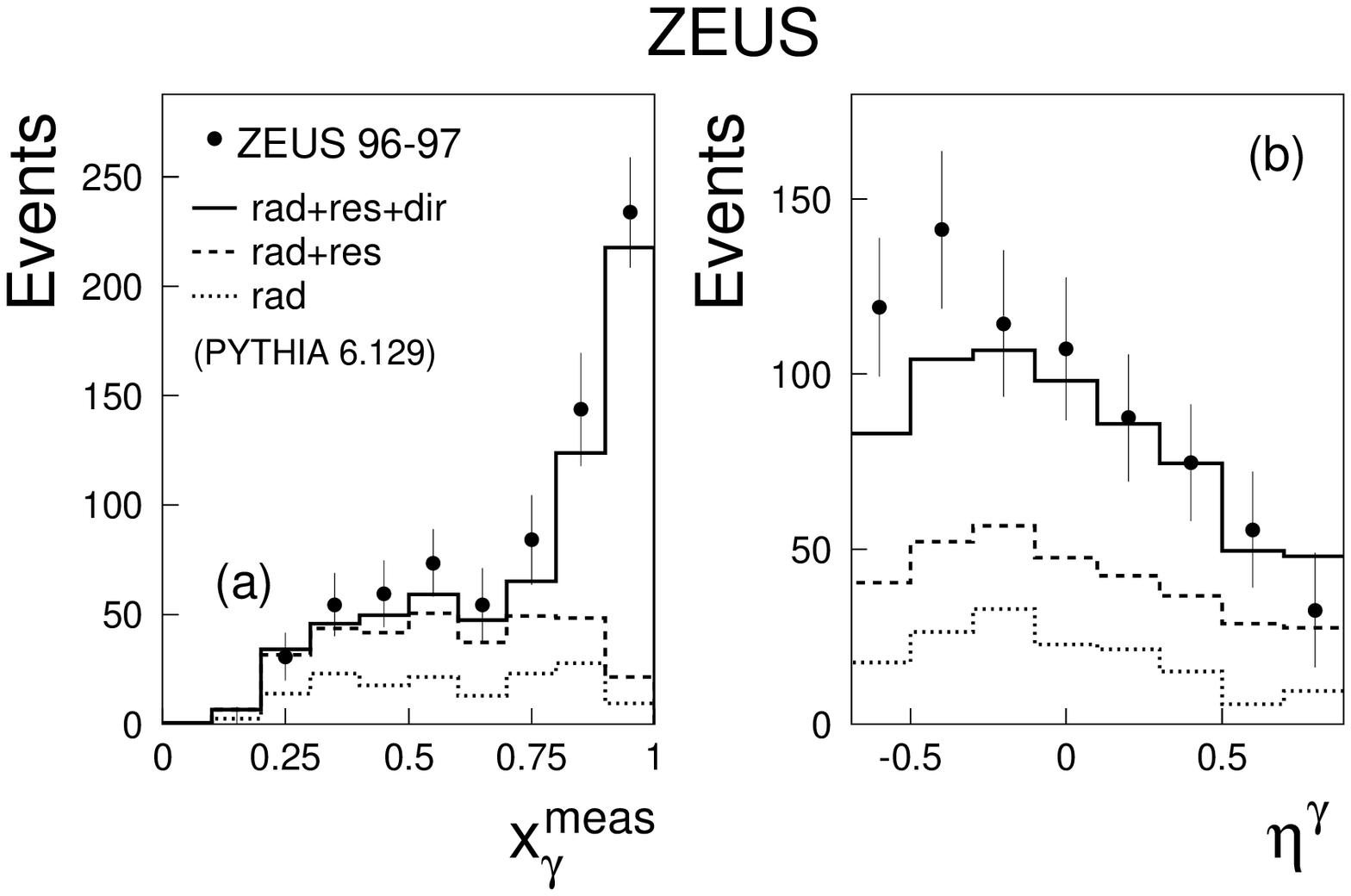, width=10.5cm}
 \epsfig{file=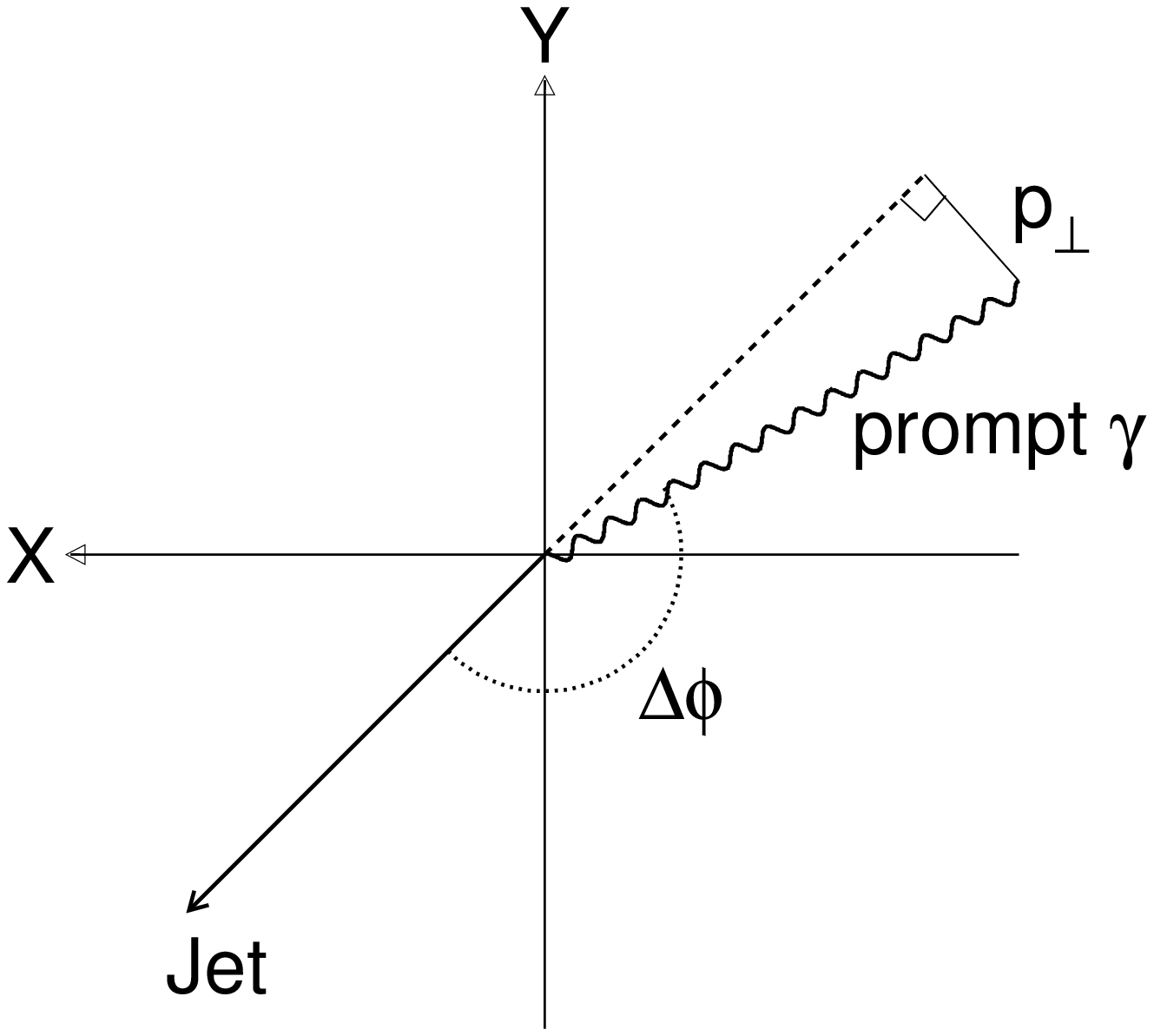, width=7cm}
\hspace*{-18mm}\raisebox{21mm}{{\small\sf(c)}}}
\caption{\small Distributions in (a) $x_\gamma^\meas$ and (b) $\eta^\gamma$ when a
jet and prompt photon are demanded, compared with PYTHIA predictions
using the GRV photon PDF, for events in the range $134 < W < 251$
GeV. (c) Kinematics of prompt photon relative to jet. }
\end{figure}

\section{Prompt photons plus jets.}
 
Figure 2 shows the distributions in $x_\gamma^\meas$ and
$\eta^\gamma$ obtained when a jet is demanded.  
The strong peak near unity in  $x_\gamma^\meas$ is due to events in which the
``direct photoproduction'' process dominates.  
In photoproduction at leading order, the Compton process
$\gamma q \to \gamma q$ is the only direct prompt photon process.
The distributions are well described by PYTHIA (c.f.\ Fig.\ 1a), giving no
suggestion for large errors in the photon PDFs.

\begin{figure}[t]
\centerline{
\epsfig{file=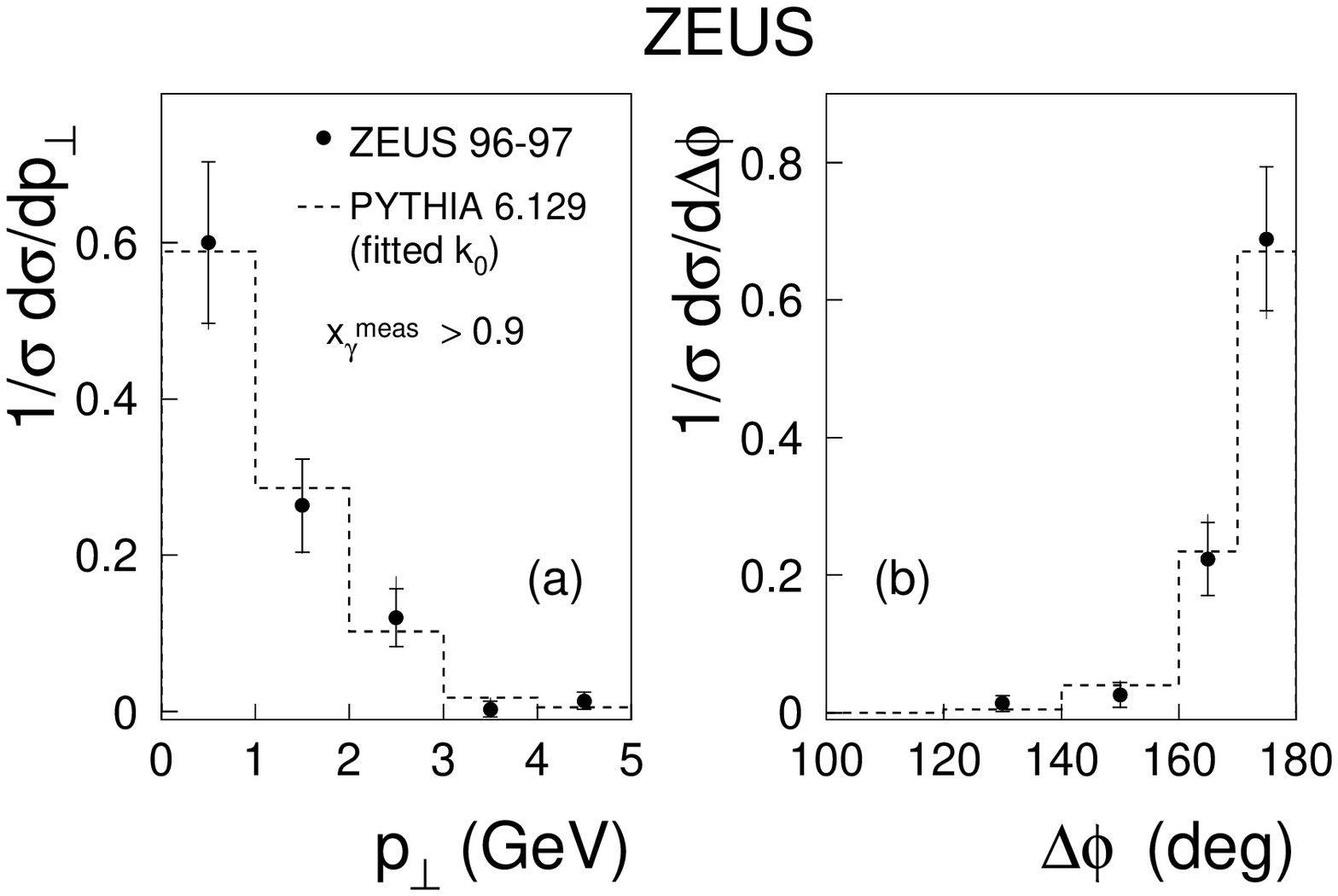,width=9.8cm,%
bbllx=33pt,bblly=385pt,bburx=520pt,bbury=710pt,clip=yes} }
\centerline{
\epsfig{file=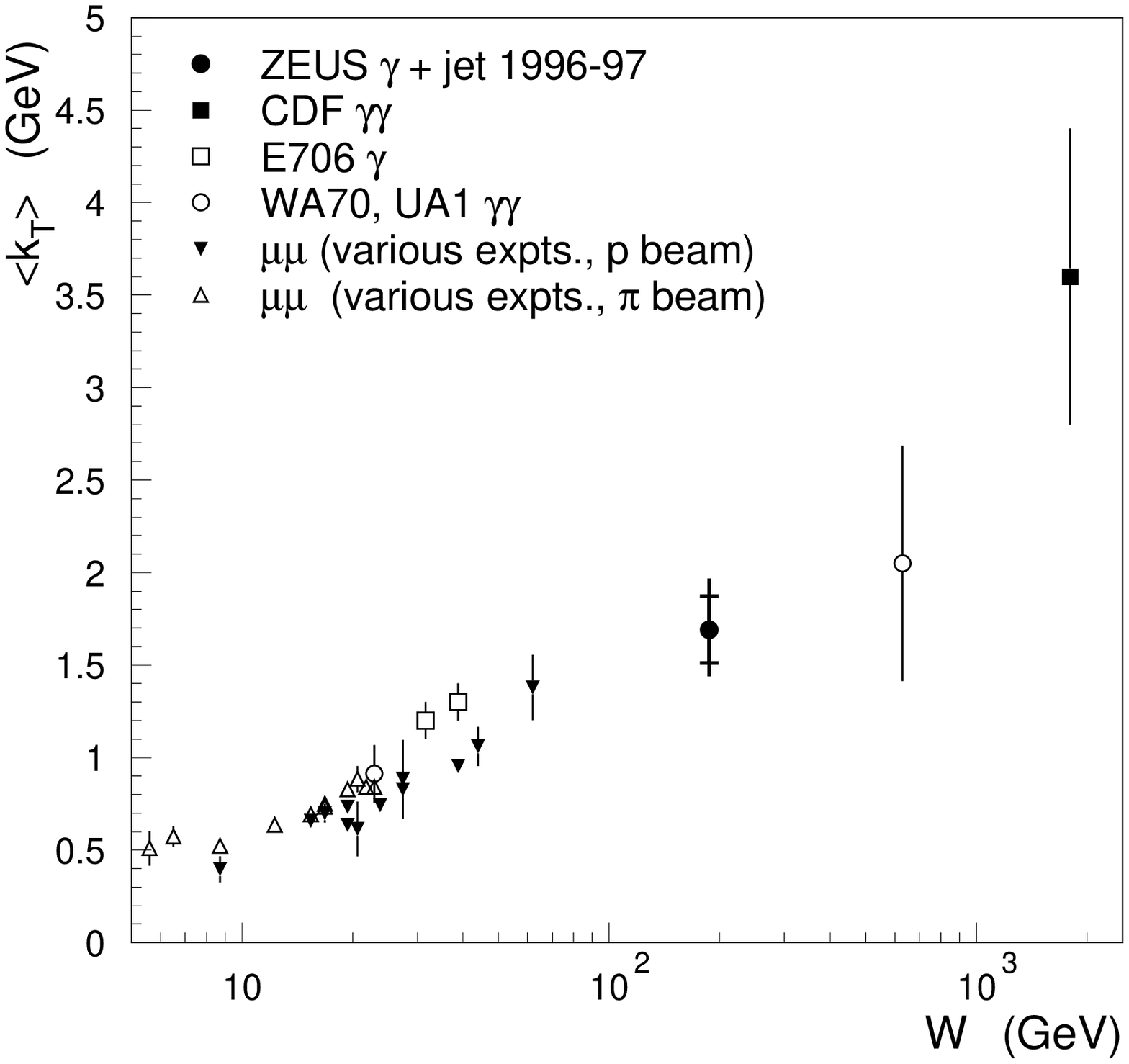,width=8cm,%
bbllx=0pt,bblly=60pt,bburx=485pt,bbury=520pt,clip=yes} 
\hspace*{-13mm}\raisebox{20mm}{{\small\sf(c)}}\hspace*{15mm}}
\caption{\small Normalised distributions of \ppp\ and $\Delta\phi$ of photon
relative to jet direction, together with fitted predictions from
PYTHIA using optimised value of $k_0$ (a,b). (c) Effective \kt\ value
in photoproduction in ZEUS, compared with results from other
experiments. (See ref.\ 2  and further references therein.)  }
\end{figure}

The condition $x_\gamma^\meas>0.9$ selects events due mainly to the
direct Compton process, and little affected by the incoming photon's
hadronic structure.  In these events, the transverse momentum \ppp\ of
the prompt photon perpendicular to the jet direction in the $r\phi$
plane is sensitive to the momentum of the quarks in the proton, as is
the azimuthal angle between the prompt photon and the jet (cf.\ Fig.\ 2c).  The
normalised distribution of \ppp\ was used to fit the PYTHIA parameter
$k_0$, the so-called ``intrinsic'' parton momentum in the proton.  A
good fit is obtained (Fig.\ 3).  By combining $k_0$ with the effects
of the initial-state parton showers, it is possible to obtain an
overall value for the effective transverse momentum of the partons in
the proton.  The result is $ \kt = 1.69\pm0.18\; ^{+0.18}_{-0.20}
\mbox{ GeV}$, as shown in Fig.\ 3c in comparison with data from a
variety of other experiments.

The ZEUS result is consistent with the trend shown by all the
measurements, namely that the value of \kt\ rises with hadronic interaction
energy.  This may be attributed to the effects of initial-state gluon
radiation, but a full theoretical description is still awaited.

\end{document}